C:/comets/Comets Page/131002

# Letter

## "The Impending Demise of Comet C/2012 S1 ISON"

**Ignacio Ferrín,**
**Institute of Physics,**
**Faculty of Exact and Natural Sciences,**
**University of Antioquia,**
**Medellin, Colombia, 05001000**
**ferrin@fisica.udea.edu.co**

**Number of pages     14**

**Number of Figures   9**

**Number of Tables    0**

## Abstract

We present evidence to conclude that comet C/2012 S1 ISON is about to turn off or disintegrate.

# 1. THE SECULAR LIGHT CURVES OF COMETS

For some years we have been developing the concept of Secular Light Curves of Comets (SLCs) (see References 1-11), a scientific way to show the brightness history of a comet. The SLCs are presented in two phase spaces, the reduced magnitude vs log of the Sun's distance, R, and the reduced magnitude vs time. Reduced means that the comet-Earth distance has been removed and only the dependence on the distance to the Sun remains. As an example, in Figure 1 we show the SLC of the famous comet 1P/Halley.

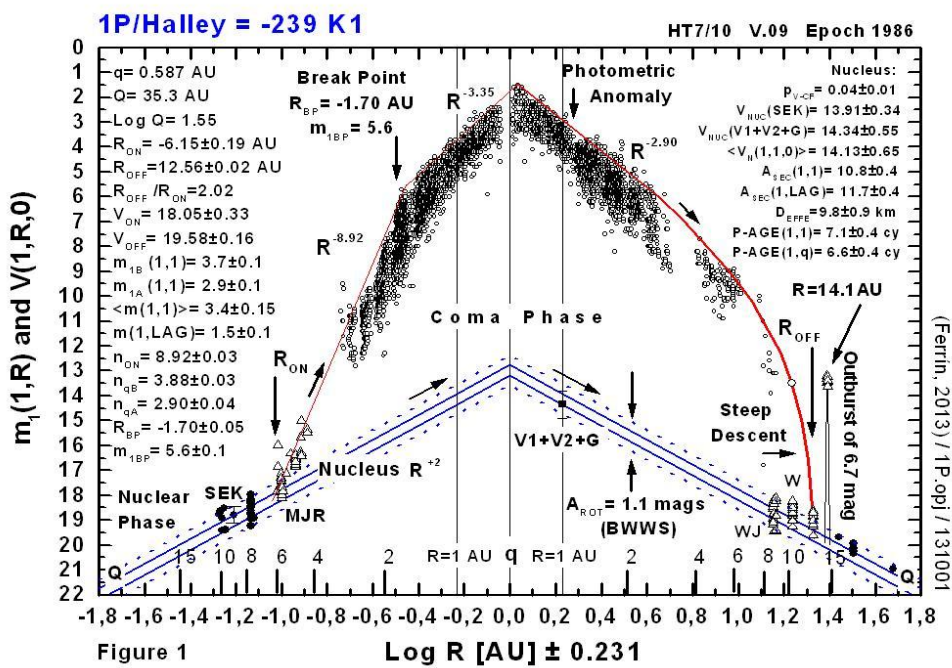

**Figure 1.** The SLC of comet 1P/Halley in the reduce magnitude vs log R phase space. This plot is updated from Reference 7.

The vertical axis is the reduced magnitude. The horizontal axis is the Log of the solar distance. Time goes from left to right but not linearly.  The line at the bottom in the form of a pyramid is the bare inactive nucleus behaving as $R^{-2}$.

Except when it is otherwise stated, in this work we adopted the *envelope of the data set* as the correct interpretation of the observed brightness.    There are many physical effects that affect comet observations like twilight, moon light, haze, cirrus clouds, dirty optics, lack of dark adaptation, excess magnification, and in the case of CCDs, sky background too bright, insufficient time exposure, insufficient CCD aperture error, and too large a scale.   All these effects diminish the captured photons coming from the comet, and the observer makes an error downward, toward fainter magnitudes. There are no corresponding physical effects that could increase the perceived brightness of a comet.   Thus the *envelope* is the correct interpretation of the data.    In fact the envelope is rather sharp, while the anti-envelope is diffuse and irregular.

We learn a lot from this plot.  There are about ~30 parameters listed, of which about ~20 are new and can be measured from the plots.  For example we learn that comets turn on and turn off.   In this case comet 1P/Halley turned on at R= -17.3 AU and turned off at +33.9 AU from the Sun.    We also learn that some comets exhibit a Slope Discontinuity Event (SDE) before perihelion that slows down the brightness rate to a more relaxed pace.    For comet 1P/Halley this took place at R= -1.70 AU.   We also notice that after the SDE the comet continued increasing in brightness steadily, at a rate $R^{-3.35}$. Now let us look at the time plot for this comet shown in Figure 2.

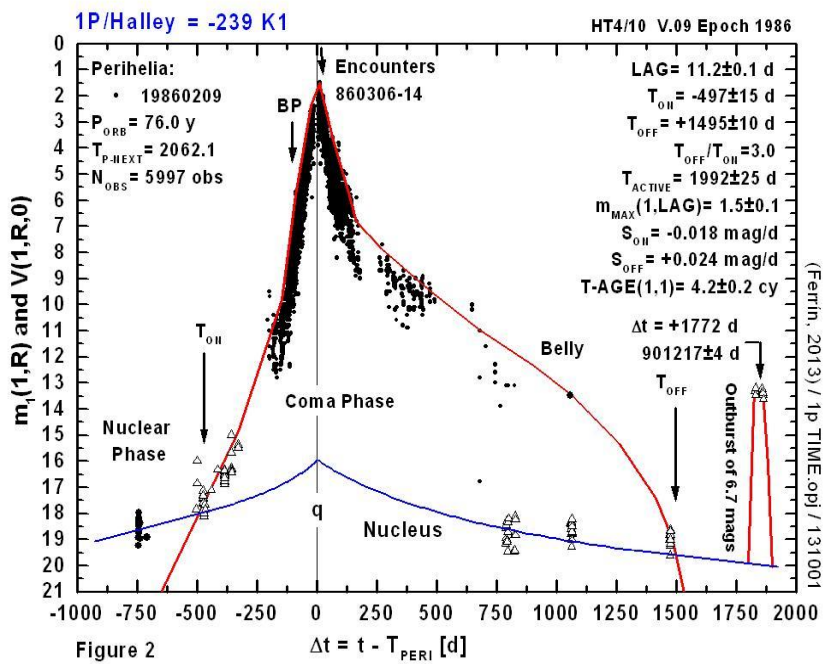

**Figure 2.** *The SLC of comet 1P/Halley in the reduced magnitude vs time phase space.*

In this plot time flows linearly from left to right. We also learn a lot from this plot. The comet exhibits a prominent belly due to the thermal wave that penetrates inside the nucleus and produces sublimation in depth. Layers inside the nucleus contain sufficient volatiles, and the nucleus continues sublimating in spite of the SDE. The plot also shows an outburst of 6.7 magnitudes of amplitude 1772 days after perihelion.

26 other SLCs appear in Reference 7, the *Atlas of Secular Light Curves of Comets, Version I.* A full interpretation of the SLCs is given there and is beyond the objectives of this paper.

However, after looking at the plots it is easy to conclude that the SLCs exhibit complexity beyond current scientific understanding.

In this investigation we reduced 11844 photometric observations of four comets.

## 2. THE SECULAR LIGHT CURVE OF COMET C/2012 S1 ISON

We are interested in creating the SLC of comet ISON to compare it with other comets.  There are several databases in the internet that give magnitudes for this comet.   We will not try to reconcile this different observation.   The easiest thing to do is to analyze each database separately.

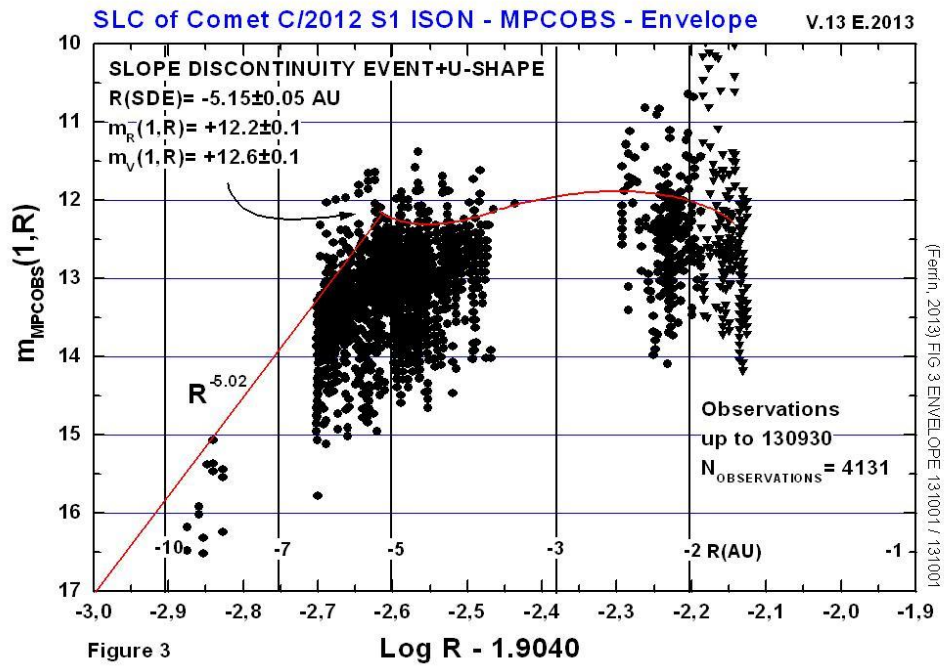

**Figure 3.**  *The SLC of comet ISON from the Minor Planet Center database (Reference 13).*

Figure 3 shows the very odd SLC of comet ISON using the Minor Planet Center database.   Three things are apparent.  First, the SDE  is very clear at a distance of around -5.1 AU pre-perihelion.   Second, there is a slight deep in the light curve just after the event with a U-shape.  And third, farther out the light curve flattens out.    If it flattens out the comet cannot be bright near perihelion.  This is an indication of trouble ahead for the comet. This plot contains observations published up to 2013 Sep 30th.

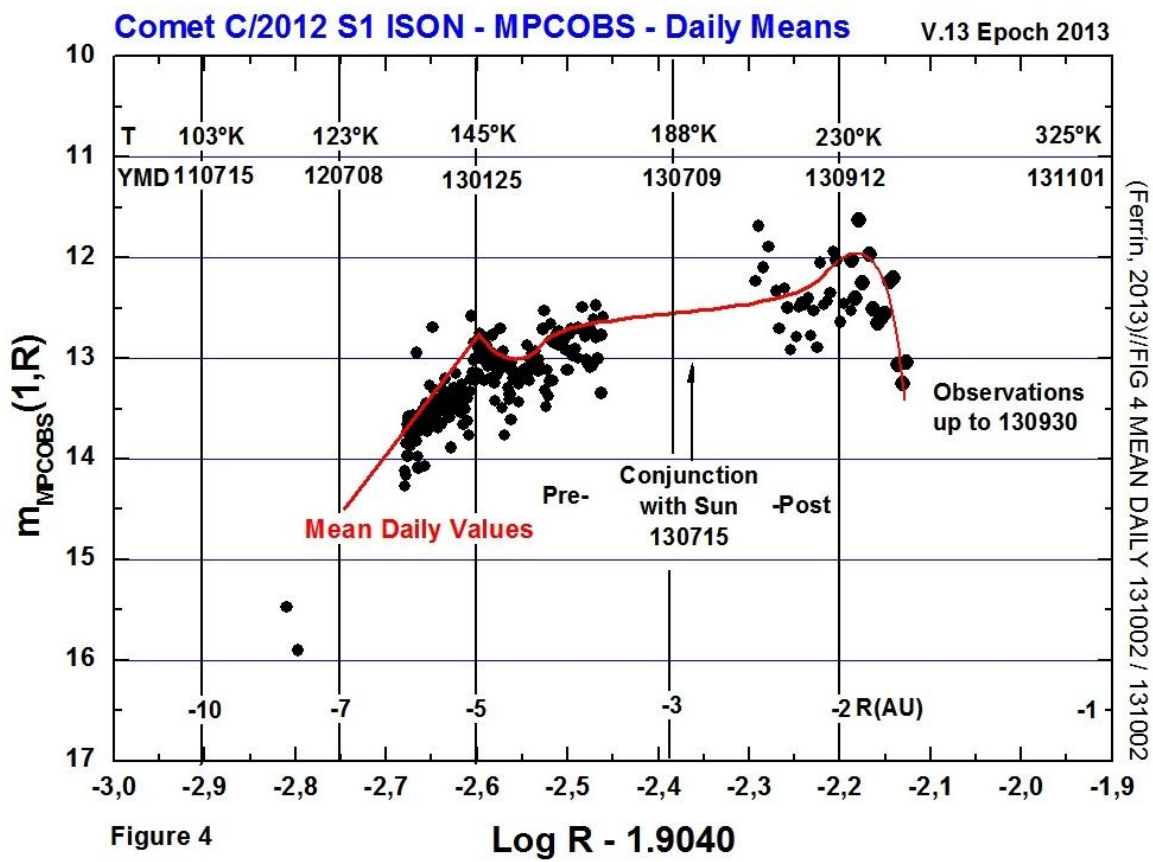

**Figure 4.**  The Minor Planet Center database is averaged daily.

To try to diminish the vertical scatter in the data we took daily mean values.  This is shown in Figure 4.  The vertical scatter has been reduced somewhat but the same trends is shown.  First, the SDE is clearly seen.  Then, there is a deep after the SDE in a U-shape.   And next, the curve flattened out.   The last 3 days, 2013 September 28th, 29th and 30th are especially worrisome because they show that the comet is precipitously decreasing its brightness by ~1 mag.

the plot the temperature of the comet is shown using a recent calibration of the temperature of a comet from Reference 9, T = 323°k / SQRT(R). It went from 103°K on 2011 Nov 15 to 230°K on 2013 Sep 12.

The temperature more than doubled but the comet ignores this fact. It is obvious that the comet is not responding to the outside energy.

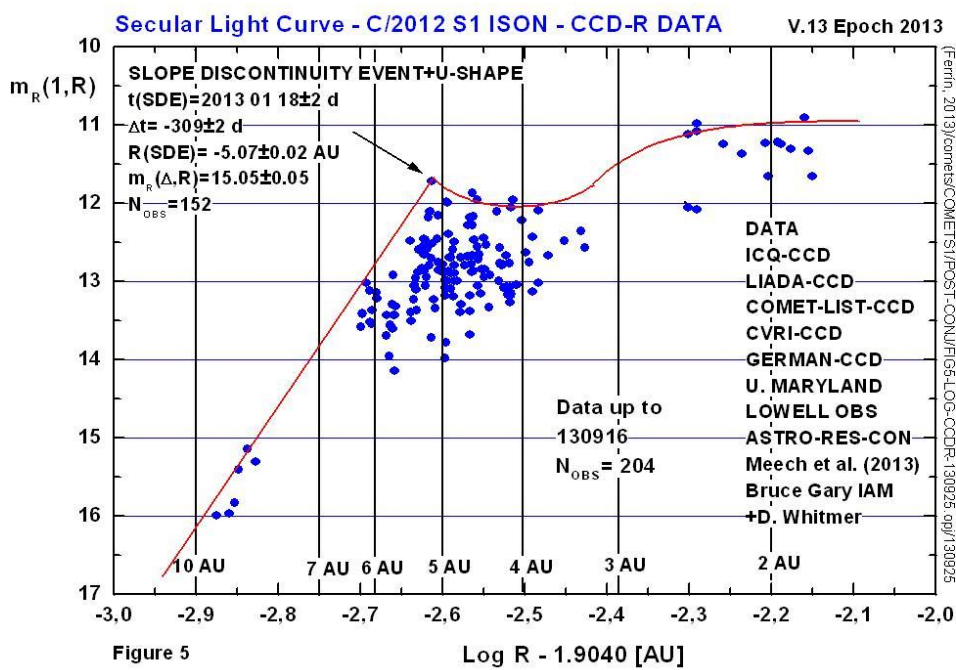

**Figure 5.** *CCD R magnitudes have been collected from the listed sites and plotted in the SLC with the same format as before.*

Next let us analyze another data set in Figure 5. This Figure plots the CCD R magnitudes collected from the literature. It shows the SDE, the U-shape of the light curve, and once again the leveling off up to 2013 September 16th. 6.7 months have passed from the SDE and the comet does not brighten significantly.

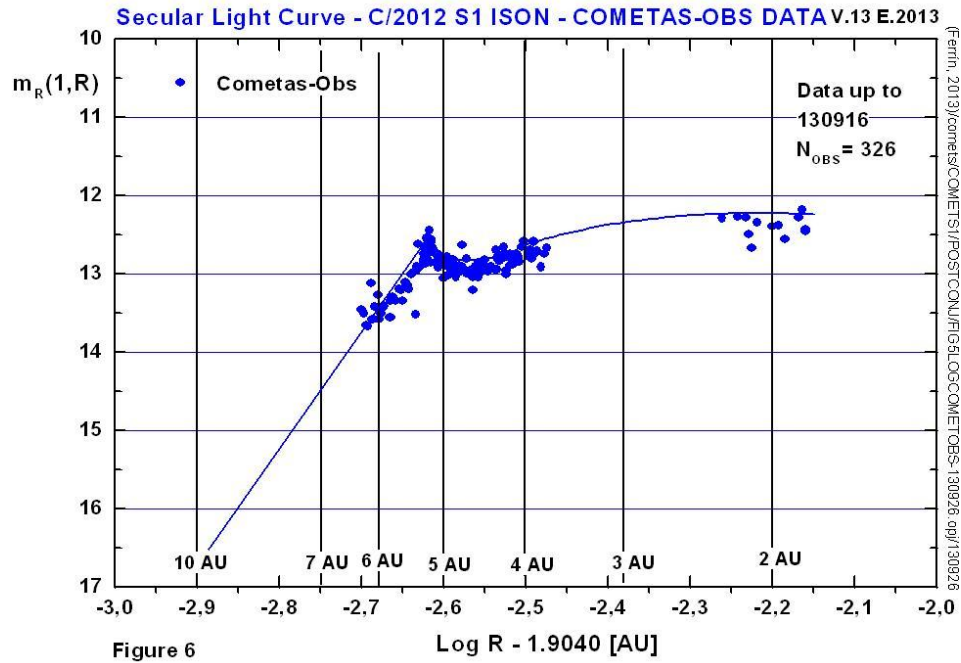

Figure 6

**Figure 6.** *The data from the Spanish observers (Reference 14).*

Finally let us look at the database from the Spanish observers, Figure 6.   It shows <u>independently</u> the same behavior exhibited by the previous plots: A SDE, a U-shape deep, and a flattened out light curve. There is no contradiction between the data sets.

There is no escape from the conclusion:  Comet ISON is not brightening at all.   In fact according to Figure 4, it has begun to fade.

Having accepted this evidence, I began a search in our database of  87 secular light curves  being prepared for the incoming ***ATLAS, Version II.***     Two instances were found with similar behavior.   The first one is comet C/2002 O4 Hönig.    The light curve is presented in Figure 7.

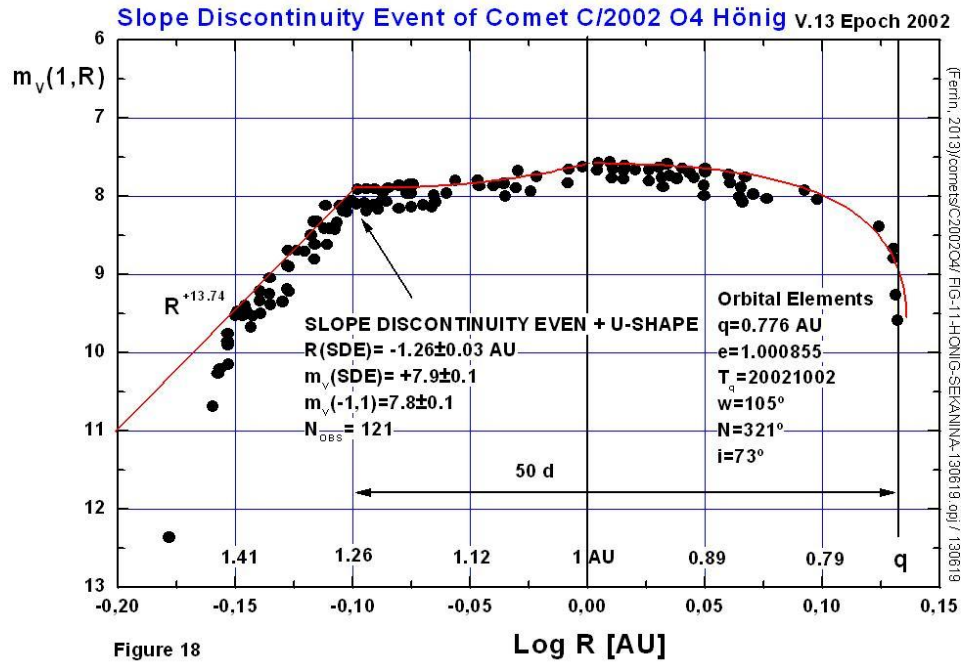

**Figure 7.** *The SLC of comet C/2002 O4 Hönig. The data for this plot comes from a paper by Sekanina (Reference 12).*

Figure 7 shows a behavior reminiscent of comet ISON. We see a clear slope discontinuity event, a slight deep in the light curve, an increase, and then the comet disintegrates after only 50 days.

A second case is found in the database, that of comet C/1996 Q1 Tabur. The SLC of this comet can be seen in Figure 8.

Figure 8 shows the same behavior followed by comet Hönig: A well-defined SDE, a U-shape deep after the event, a leveling off and then a precipitous decay into disintegration.

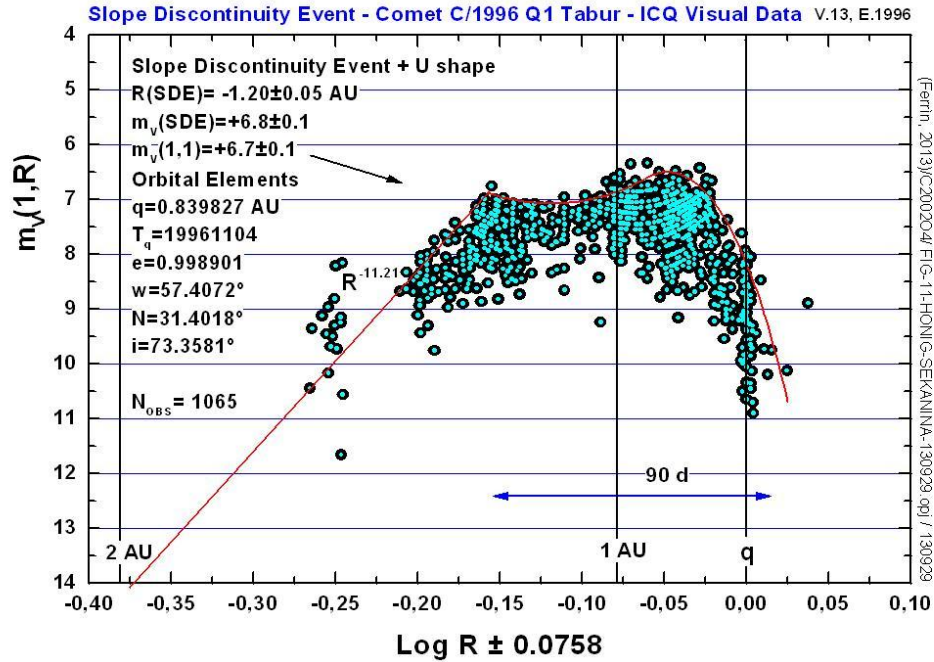

**Figure 8.** *The SLC of comet C/1996 Q1 Tabur. The data set for this plot comes from the ICQ database maintained by Daniel Green (Reference 15) .*

## 3. WHAT IS THE PROBABILITY OF COMET ISON TURNING OFF OR DISINTEGRATING ?

In view of the evidence  presented above there is a 100% probability that comet ISON is  turning off or disintegrating.  The reason is that it exhibits the same SDE+U-shape signature as comets Hönig and Tabur that disintegrated.

This dispels the notion that comets are not predictable.

Comets announce that they are going to disintegrate by exhibiting the SDE+U-shape signature.

Also, the very sharp discontinuity of slope at the SDE implies that this cannot be an outburst. There is some fundamental physical process that goes on here that we do not understand. Or if this is an outburst, it is a different kind of outburst than the ones we have seen up to now.

The location of the Slope Discontinuity Event is also puzzling in the following diagram, Figure 9.

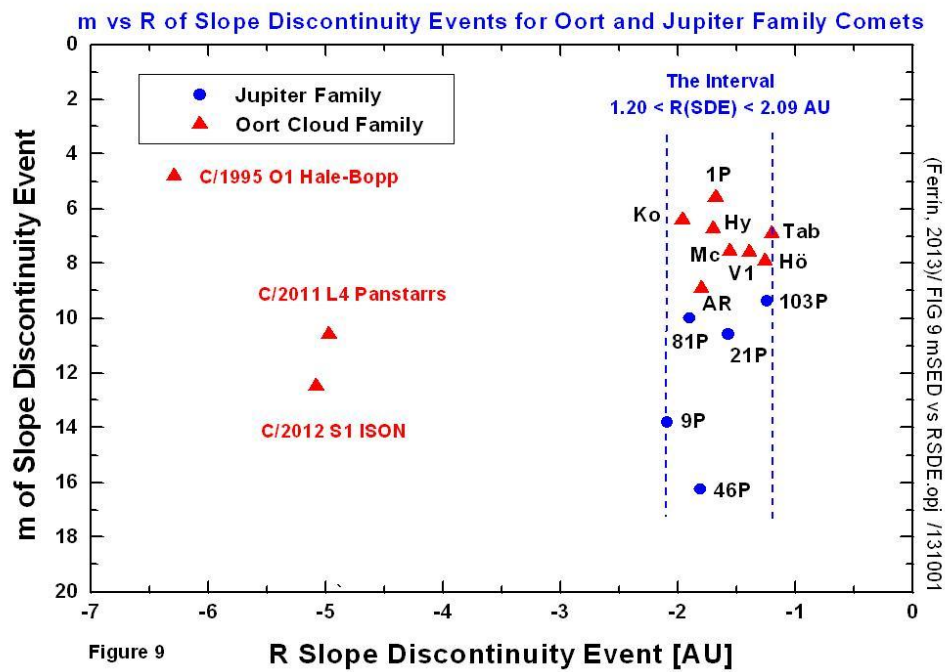

**Figure 9.** *The location of the SDE is plotted vs solar distance. Notice how close comet Hönig is to comet Tabur and how far they are from comet ISON.*

Figure 9 shows that the SDEs are located either far from the Sun or in a narrow interval from 1.20 to 2.09 AU. What is surprising here is that the same signature appears at so different distances.  This is telling us that the signature is distance independent.

## 4. CONCLUSIONS

a. In the last three days of September the comet has decreased its brightness by a large amount, ~1.0 magnitudes suggesting that the turn off has begun.

b. The temperature doubled in the observed interval and the comet is not responding.

c. The comet passed the "frost line" set by some astronomers at 2.5 to 3 AU and nothing happened.

d. The comet exhibits a Slope Discontinuity Event + U-Shape previously exhibited by two disintegrating comets: Hönig and Tabur.

e. The SLCs exhibit complexity beyond current scientific understanding.

The conclusion is simple: comet ISON has begun to turn off or disintegrate.

We had advanced this hypothesis on June 19[th], 2013 (Reference 11).

Next few days and weeks are crucial to determine the fate of this object.

## 5. NOTICE

Due to the nature of this investigation you are free to use the information provided in this report.  However we request that scientific ethics be followed and that the source be quoted, in accord with good scientific practices.

**REFERENCES**
1. **Ferrín, I., 2005a. Variable Aperture Correction Method in Cometary Photometry, ICQ 27, p. 249-255.**
2. **Ferrín, I., 2005b. "Secular Light Curve of Comet 28P/Neujmin 1, and of Comets Targets of Spacecraft, 1P/Halley, 9P/Tempel 1, 19P/Borrelly, 21P/Grigg-Skejellerup, 26P/Giacobinni-Zinner, 67P/Chruyumov-Gersimenko, 81P/Wild 2". Icarus 178, 493-516.**
3. **Ferrín, I., 2006. "Secular Light Curve of Comets: 133P/Elst-Pizarro". Icarus, 185, 523-543.**
4. **Ferrín, I., 2007. "Secular Light Curve of Comet 9P/Tempel 1". Icarus, 187, 326-331.**
5. **Ferrín, I., 2008. "Secular Light Curve of Comet 2P/Encke, a comet active at aphelion". Icarus, 197, 169-182.**
   http://arxiv.org/ftp/arxiv/papers/0806/0806.2161.pdf
6. **Ferrín, I., 2009. "Secular Light Curve of Comet 103P/Hatley 2, the next target of the Deep Impact EPOXI Mission". PSS, 58, 1868-1879.**
   http://arxiv.org/ftp/arxiv/papers/1008/1008.4556.pdf
7. **Ferrín, I., 2010. "Atlas of Secular Light Curves of Comets". PSS, 58, 365-391.**
   http://arxiv.org/ftp/arxiv/papers/0909/0909.3498.pdf
8. **Ferrín, I., Hamanowa, H., , Hamanowa, H., Hernández, J., Sira, E., Sánchez, A., Zhao, H., Miles, R., 2012. "The 2009 apparition of methuselah comet 107P/Wilson-Harrington: A case of comet rejuvenation?". PSS, 70, 59-72.**
   http://arxiv.org/ftp/arxiv/papers/1205/1205.6874.pdf
9. **Ferrín, I., Zuluaga, J., Cuartas, P., 2013. "The location of Asteroidal Belt Comets on a Comets' evolutionary diagram: The Lazarus Comets". MNRAS in press.**
   http://arxiv.org/ftp/arxiv/papers/1305/1305.2621.pdf
10. **Ferrín, I. 2013. "The secular light curves of comets C/2011 L4 and C/2012 S1 compared to that of comet 1P/Halley".**
   http://arxiv.org/ftp/arxiv/papers/1302/1302.4621.pdf
11. **Ferrín, I. "The Location of Oort Cloud Comets C/2011 L4 Panstarrs and C/2012 S1 ISON, on a Comets' Evolutionary Diagram".**
   http://arxiv.org/ftp/arxiv/papers/1306/1306.5010.pdf
12. **Sekanina, Z., 2002. ICQ, 24, 223-236.**
13. **Minor Planet Center repository of astrometric observations, http://www.minorplanetcenter.net/db_search**

14. The spanish group measures magnitudes with several CCD
    apertures:
      **http://www.astrosurf.com/cometas-obs/** .  It is managed by Julio
      Castellanos,  Esteban Reina and Ramon Naves.
15. The Cometary Science Archive
      **http://www.csc.eps.harvard.edu/index.html**
      is a site to visit because it contains useful scientific information
      of current and past comets, and it is maintained by Daniel
      Green.